\documentclass{amia}
\usepackage{lipsum}

\usepackage{amsmath}    
\usepackage{amssymb}    
\usepackage{mathtools}  
\usepackage{amsfonts}   
\usepackage{amsthm}
\usepackage{xcolor}

\usepackage{booktabs}   
\usepackage{longtable}  
\usepackage{tabularx}   
\usepackage{array}

\usepackage{graphicx}
\usepackage{float}
\usepackage{makecell}
\usepackage{wrapfig}
\usepackage{needspace}
\usepackage[super,sort&compress,comma]{natbib}
\usepackage{tikz}
\tikzset{
    topnum/.style   ={font=\rmfamily\fontsize{14}{14}\selectfont},
    topname/.style  ={font=\rmfamily\fontsize{14}{16}\selectfont},
    toplabel/.style ={font=\rmfamily\fontsize{12}{12}\selectfont},
    toptext/.style  ={font=\rmfamily\fontsize{12}{14}\selectfont, align=center},
    topformula/.style={font=\rmfamily\fontsize{13}{15}\selectfont},
    toptime/.style  ={font=\rmfamily\itshape\fontsize{16}{16}\selectfont},
    eventfont/.style={font=\bfseries\fontsize{10}{10}\selectfont},
    epochfont/.style={font=\bfseries\fontsize{10}{10}\selectfont},
    timelab/.style  ={font=\fontsize{13}{13}\selectfont}
}
\usetikzlibrary{arrows.meta, positioning, calc, shapes.geometric, shadows, backgrounds, fit, decorations.pathreplacing, decorations.pathmorphing}

\newcommand{\event}[1]{\mathsf{#1}}

\newcommand{\pEvent}[1]{\varphi_{\event{#1}}}

\theoremstyle{definition}

\newcommand*{\semantics}[1]{\textnormal{[\kern-.15em[}#1\textnormal{]\kern-.15em]}}

\makeatletter
\renewcommand\section{\@startsection {section}{1}{\z@}%
                     {-0.45\baselineskip} {0.2\baselineskip} 
                     {\normalfont\fontsize{10pt}{10pt}\bfseries\rmfamily}}
\renewcommand\subsection{\@startsection {subsection}{1}{\z@}%
                     {-0.75\baselineskip} {0.2\baselineskip} 
                     {\normalfont\fontsize{10pt}{10pt}\bfseries\rmfamily}}
\renewcommand\subsubsection{\@startsection {subsubsection}{1}{\z@}%
                     {-0.75\baselineskip} {0.2\baselineskip} 
                     {\normalfont\fontsize{10pt}{10pt}\bfseries\rmfamily}}
\makeatother
\begin{document}

\title{Ensemble Logic for Symbolic Representation of Sleep Medicine Guidelines}

\author{
Jiahao Fan, PhD$^{1,3}$,
Xiaojin Li, PhD$^{1,3}$,
Yan Huang, PhD$^{1,3}$,
Xubing Hao, PhD$^{1,3}$,\\
Licong Cui, PhD$^{2,3}$,
Guo-Qiang Zhang, PhD$^{1,2,3}$*
}

\institutes{
$^{1}$Department of Neurology, 
The University of Texas Health Science Center at Houston, Houston, Texas, 77030, USA; $^{2}$McWilliams School of Biomedical Informatics, The University of Texas Health Science Center at Houston, Houston, Texas, 77030, USA; $^{3}$Texas Institute for Restorative Neurotechnologies, The University of Texas Health Science Center at Houston, Houston, Texas, 77030, USA\\[4pt]
\textbf{*Corresponding author:} Guo-Qiang.Zhang@uth.tmc.edu
}

\maketitle

\section*{Abstract}

\textit{The American Academy of Sleep Medicine (AASM) Manual is the clinical standard for polysomnography (PSG) scoring, but its narrative rules can admit multiple reasonable interpretations, contributing to inter-scorer variability and implementation differences across studies and software systems. We present a formal framework for translating sleep-scoring rules into Rational Ensemble Logic (QEL), a dense-time (i.e., a continuous, rational-valued timeline rather than discrete steps) formalism that combines first-order quantification with metric temporal operators. Using an extraction-and-compilation procedure, we identified 18 unique atomic propositions and derived 12 final specifications corresponding to clinically scoreable AASM events. Back-translation of QEL specifications into clinician-facing language retained high semantic fidelity to the original scoring narratives (embedding cosine similarity: 79.3, 95\% CI: 79.0--79.7) despite low lexical overlap (ROUGE-L: 18.3, 95\% CI: 17.6--18.9). Formalization also clarifies latent ambiguities, including implicit physiological latencies and overlapping exclusions. This framework yields executable, rigorous rule specifications for computational phenotyping, more consistent implementation across datasets, and standardized open-source PSG analysis.}

\section*{1. Introduction}

In clinical care and sleep research, polysomnography (PSG) is the gold standard for assessing physiological signals during sleep. To standardize interpretation of these multimodal signals, the \emph{American Academy of Sleep Medicine} (AASM) Scoring Manual \cite{berry2017aasm_update,berry2017aasm} defines clinical guidelines for sleep staging and event scoring. These guidelines function as a ``universal common language'' across sleep medicine. However, free-text rules lack explicit mathematical structure and formal constraints. Many sleep events involve linking changes across signals, enforcing minimum durations, and applying exclusions over time; when these constraints remain in prose, different implementations may operationalize the same guideline differently. This creates a semantic gap between phenotype definitions and executable logic, limiting rigorous reuse of large-scale real-world data (RWD) for cohort identification, phenotyping, and clinical research, and motivates translating narrative rules into an executable logical formalism.

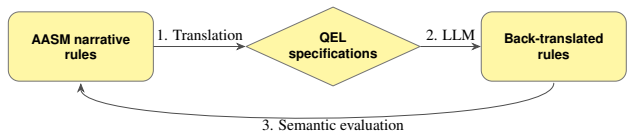
\begin{wrapfigure}{l}{0.5\textwidth}
    \centering
    \vspace{-6pt}
    \resizebox{\linewidth}{!}{
    \begin{tikzpicture}[
        >=Stealth,
        line cap=round,
        line join=round,
        font=\sffamily,
        block/.style={
            draw=black!55,
            thick,
            rounded corners=10pt,
            minimum width=4.6cm,
            minimum height=2.2cm,
            align=center,
            font=\sffamily\bfseries\large,
            fill=orange!18!yellow!35
        },
        diamondblock/.style={
            draw=black!55,
            thick,
            diamond,
            aspect=2.2,
            minimum width=4.3cm,
            minimum height=2.3cm,
            font=\sffamily\bfseries\large,
            align=center,
            fill=orange!18!yellow!35
        },
        flow/.style={
            draw=black!70,
            thick,
            -{Stealth[length=3mm,width=2.4mm]}
        },
        evalflow/.style={
            draw=black!70,
            thick,
            -{Stealth[length=3mm,width=2.4mm]}
        }
    ]

    \node[block] (src) at (-8.0,0)
        {AASM narrative\\rules};

    \node[diamondblock] (logic) at (0,0)
        {QEL\\specifications};

    \node[block] (tgt) at (7,0)
        {Back-translated\\rules};

  \draw[flow] (src.east) -- node[above=2pt, font=\fontsize{14pt}{10pt}\selectfont] {1. Translation} (logic.west);
    \draw[flow] (logic.east) -- node[above=2pt, font=\fontsize{14pt}{10pt}\selectfont] {2. LLM} (tgt.west);

    \coordinate (A) at ($(tgt.south)+(0,-0.05)$);
    \coordinate (B) at ($(src.south)+(0,-0.05)$);

    \draw[evalflow]
        (A)
        .. controls ($(A)+(0,-1.45)$) and ($(B)+(0,-1.45)$) ..
        (B);

    \node[font=\fontsize{14pt}{10pt}\selectfont] at (0,-2.5) {3. Semantic evaluation};

    \end{tikzpicture}
    }
    \caption{High-level overview of the study paradigm. Narrative AASM scoring rules are translated into QEL formal specifications, then back-translated into clinician-facing rule descriptions, and finally evaluated against the original source rules for semantic fidelity. }
    \label{fig:study_paradigm}
    \vspace{0.3\baselineskip}
\end{wrapfigure}

Despite the clinical indispensability of PSG, there is no practical framework for formally representing and reasoning about PSG-derived events and phenotypes. Existing temporal formalisms often face a trade-off between expressive power and computational feasibility. Linear Temporal Logic (LTL) \cite{ltl} and Metric Temporal Logic (MTL) \cite{mtl} are point-based and therefore do not natively capture interval semantics, such as durations, that are central to sleep physiology. In contrast, interval-based formalisms, such as Interval Temporal Logic (ITL) \cite{itl,stanforditl} and Halpern--Shoham (HS) logic \cite{hs}, can represent rich interval relationships. However, they incur substantial representational overhead, high computational cost \cite{hs_ultimiate}, and limited agility for precise cross-timeline, point-based reference.

To address this impasse, we use Ensemble Logic (EL), a general framework for representing temporal statements over both discrete and dense time~\cite{zhang2024temporal,kr}. Prior work showed that Temporal EL over a discrete timeline can represent complex clinical trial protocols~\cite{time-trial}. PSG, however, is naturally a dense-time record: signal samples and annotation boundaries are better understood as time points and intervals on a continuous timeline than as coarse stepwise increments. We therefore use Rational Ensemble Logic (QEL), the dense-time instantiation of EL. For PSG, QEL is particularly useful because it provides five simple components that closely mirror how sleep scoring rules are described in practice:

\begin{itemize}
    \setlength{\itemsep}{1pt}
    \setlength{\parsep}{0pt}
    \setlength{\parskip}{0pt}
    \setlength{\topsep}{2pt}
    \setlength{\partopsep}{0pt}
    \item \textbf{Shift} $(p_{x+t})$: places a finding at a specific offset from a reference time point.  
    For example, snore occurs 6 seconds later can be formalized as $\event{Snore}_{6}$.

    \item \textbf{Box} $(\Box_t p)$: states that a condition holds continuously for a duration $t$.  
    For example, airflow reduction greater than 30\% persisting for 10 seconds can be formalized as $\Box_{10}\event{FlowDrop30}$.

    \item \textbf{Diamond} $(\Diamond_t p)$: states that an event occurs at least once within a bounded window of length $t$.  
    For example, a spindle occurring within a 3-second window can be formalized as $\Diamond_3\event{Spindle}$.

    \item \textbf{Boolean connectives} $(\wedge, \vee, \neg)$: combine findings across channels or exclude competing conditions.  
    For example, an abrupt EEG event occurring during stage N2 can be formalized as $\event{EEGAbrupt} \wedge \event{N2}$.

    \item \textbf{Quantification} $(\exists x, \forall x)$: allows different parts of a rule to refer to the same time anchor or event interval.  
    For example, the existence of a K-complex at time $x$ can be formalized as $\exists x\, \event{KC}_x$.
\end{itemize}

These five components align naturally with sleep-medicine rules, which are typically expressed in terms of fixed offsets, minimum durations, bounded timing windows, concurrent multi-signal criteria, and repeated reference to the same scored epoch. They also match how PSG annotations are stored in practice: as labeled time intervals that together form a Biomedical Event Structure Temporal Model (BEST) \cite{kr}. This makes it natural to lift a continuous PSG record into an explicit symbolic model, so that complex scoring rules can be encoded as executable formulas and evaluated systematically by model checking.

In this work, we apply QEL to the AASM Scoring Manual and present a principled framework for deriving formal specifications of sleep phenotypes from narrative scoring rules (Fig. \ref{fig:study_paradigm}). Our approach translates key components of the manual from free-text descriptions into logical specifications with machine-checkable semantics for sleep scoring. This process yields a unified mathematical representation of sleep phenotypes, including apneas, hypopneas, arousals, and limb movements. Concretely, we define practical principles for compiling sleep events, temporal constraints, and logical dependencies into QEL constructs. Because QEL specifications are assembled from a compact set of interpretable components, the mapping between free-text protocol statements and logic is bidirectional. We leverage this property in a large language model (LLM)-assisted back-translation evaluation, in which QEL specifications are verbalized into clinician-facing narratives and compared with the source AASM text for semantic fidelity and lexical overlap. We provide QEL formulas for representative events spanning multiple chapters of the AASM Manual and show how formalization resolves latent semantic ambiguities in the guidelines.

\section*{2. Background}

\subsection*{2.1\quad Overview of Sleep Scoring}
PSG contains heterogeneous physiological signals to characterize sleep architecture. A standard PSG recording captures three primary categories of physiological signals: electrophysiology signals (EEG, EOG, EMG, ECG), respiratory signals (airflow, effort, oximetry), and movement parameters. The AASM standards recommend sampling rates typically ranging from 200 Hz to 500 Hz for EEG/EOG/EMG to ensure the capture of transient waveforms such as spikes and sleep spindles.

\definecolor{mainblue}{RGB}{77,110,190}
\definecolor{deepblue}{RGB}{42,63,125}
\definecolor{eventred}{RGB}{236,57,41}
\definecolor{epochorange}{RGB}{221,164,118}
\definecolor{eventgold}{RGB}{191,166,70}
\definecolor{panelgray}{RGB}{215,217,222}
\definecolor{boxgray}{RGB}{236,236,236}
\definecolor{dashgray}{RGB}{145,145,145}

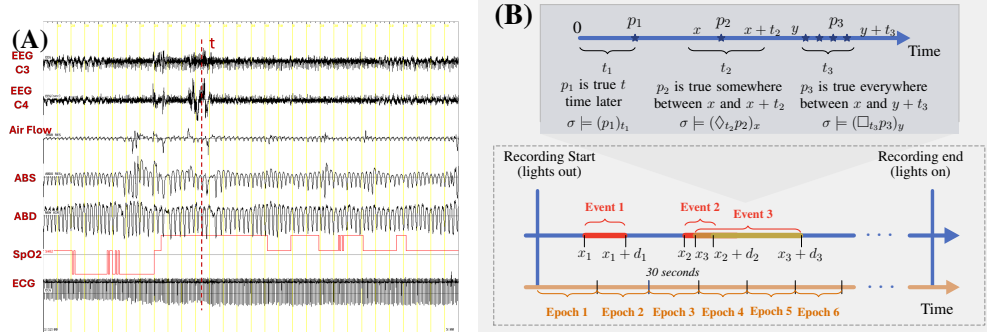
\begin{figure}[htbp]
\centering
\resizebox{0.8\linewidth}{!}{
\begin{minipage}[t]{0.8\textwidth}
    \centering
    \begin{tikzpicture}
        \node[anchor=south west, inner sep=0] (img) at (0,0)
        {\includegraphics[width=\linewidth]{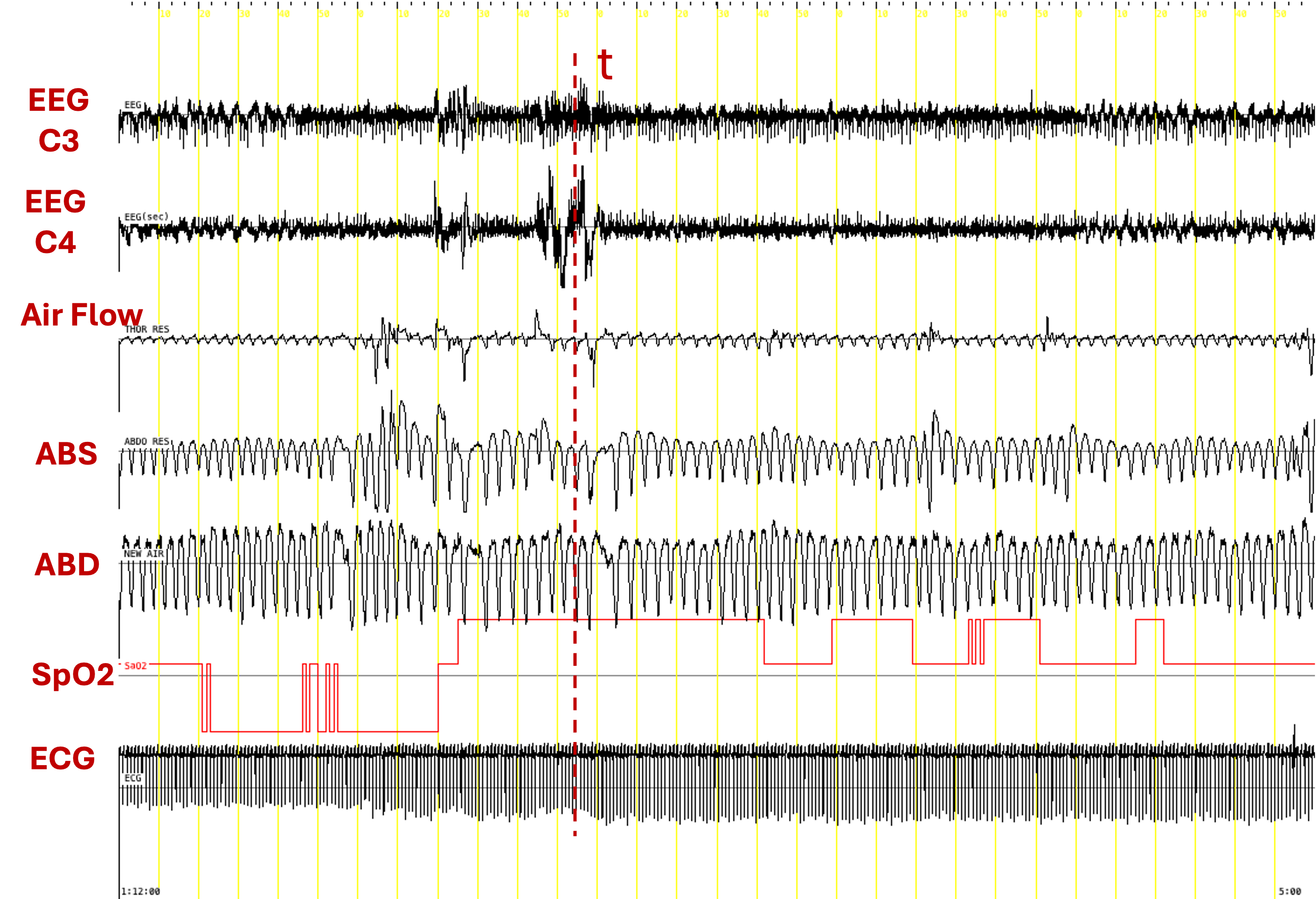}};
        
    \node[
        anchor=north west,
        font=\bfseries\fontsize{22}{22}\selectfont
    ] at ([xshift=4pt,yshift=-4pt]img.north west) {(A)};
    \end{tikzpicture}
\end{minipage}
\hspace{5mm}%
\begin{tikzpicture}[
    x=1cm,y=1cm,
    line cap=round,
    line join=round,
    >=Stealth
]
\hfill
\begin{minipage}[c]{0.49\textwidth}
    \centering
\tikzset{
    topnum/.style   ={font=\rmfamily\fontsize{14}{14}\selectfont},
    topname/.style  ={font=\rmfamily\fontsize{14}{16}\selectfont},
    toplabel/.style ={font=\rmfamily\fontsize{12}{12}\selectfont},
    toptext/.style  ={font=\rmfamily\fontsize{12}{14}\selectfont, align=center},
    topformula/.style={font=\rmfamily\fontsize{13}{15}\selectfont},
    toptime/.style  ={font=\rmfamily\itshape\fontsize{16}{16}\selectfont},
    eventfont/.style={font=\bfseries\fontsize{10}{10}\selectfont},
    epochfont/.style={font=\bfseries\fontsize{10}{10}\selectfont},
    timelab/.style  ={font=\fontsize{13}{13}\selectfont}
}
\end{minipage}
\fill[gray!12] (0,-7.75) rectangle (15,2.0);

\node[anchor=north west,font=\bfseries\fontsize{24}{24}\selectfont]
    at (0.35,1.9) {(B)};

\fill[panelgray] (1.8,-2) rectangle (13.95,1.75);
\draw[gray!55,line width=.4pt] (1.8,1.75) -- (13.95,1.75);   
\draw[gray!55,line width=.4pt] (1.8,-1.65) -- (1.8,1.75);    
\draw[gray!55,line width=.4pt] (13.95,-1.65) -- (13.95,1.75);

\draw[mainblue,line width=1.2mm,-{Triangle[length=4mm,width=4mm]}]
    (2.95,0.78) -- (12.55,0.78);

\node[topnum]  at (2.88,1.08) {0};
\node[topname] at (4.55,1.23) {$p_1$};
\node[topname] at (7.12,1.23) {$p_2$};
\node[topname] at (10.40,1.23) {$p_3$};

\node[anchor=west,font=\fontsize{13}{13}\selectfont]
    at (12.35,0.40) {Time};

\node[star,star points=5,star point ratio=2.25,
      fill=deepblue,draw=deepblue,inner sep=1.0pt] at (4.55,0.78) {};
\node[star,star points=5,star point ratio=2.25,
      fill=deepblue,draw=deepblue,inner sep=1.0pt] at (7.05,0.78) {};

\foreach \xx in {9.50,9.90,10.30,10.70}{
    \node[star,star points=5,star point ratio=2.25,
          fill=deepblue,draw=deepblue,inner sep=1.0pt] at (\xx,0.78) {};
}

\draw[decorate,decoration={brace,amplitude=5pt,mirror},line width=.8pt]
    (2.95,0.5) -- (4.55,0.5)
    node[midway,below=9pt,toplabel] {$t_1$};

\draw[decorate,decoration={brace,amplitude=5pt,mirror},line width=.8pt]
    (6.10,0.5) -- (8.30,0.5)
    node[midway,below=9pt,toplabel] {$t_2$};

\draw[decorate,decoration={brace,amplitude=5pt,mirror},line width=.8pt]
    (9.40,0.5) -- (10.90,0.5)
    node[midway,below=9pt,toplabel] {$t_3$};

\node[toplabel] at (6.35,1.0) {$x$};
\node[toplabel] at (8.25,1.0) {$x+t_2$};
\node[toplabel] at (9.2,1.0) {$y$};
\node[toplabel] at (11.6,1.0) {$y+t_3$};

\node[toptext, text width=3.5 cm] at (3.35,-0.8)
{$p_1$ is true $t$ time later};

\node[toptext, text width=5cm] at (7.00,-0.90)
{$p_2$ is true somewhere\\between $x$ and $x+t_2$};

\node[toptext, text width=3.8cm] at (11.20,-0.90)
{$p_3$ is true everywhere\\between $x$ and $y+t_3$};

\node[topformula] at (3.5,-1.7)
{$\sigma \models (p_1)_{t_1}$};

\node[topformula] at (7.02,-1.7)
{$\sigma \models (\Diamond_{t_2}p_2)_x$};

\node[topformula] at (11.18,-1.7)
{$\sigma \models (\Box_{t_3}p_3)_y$};

\fill[gray!18] (1.8,-2) -- (13.8,-2) -- (7.75,-4) -- cycle;

\draw[
    dash pattern=on 3pt off 2pt,
    color=dashgray,
    line width=.45pt
] (0.45,-7.55) rectangle (14.70,-2.35);

\node[
    anchor=west,
    align=center,
    font=\fontsize{12}{12}\selectfont
] at (0.62,-2.9)
{Recording Start\\(lights out)};

\node[
    anchor=east,
    align=center,
    font=\fontsize{12}{12}\selectfont
] at (14.20,-2.9)
{Recording end\\(lights on)};

\draw[mainblue,line width=1.0mm] (1.72,-6.45) -- (1.72,-3.65);
\draw[mainblue,line width=1.0mm] (13.18,-6.45) -- (13.18,-3.65);

\draw[mainblue,line width=1.2mm]
    (1.38,-4.95) -- (10.90,-4.95);
\draw[mainblue,line width=1.2mm,-{Triangle[length=4.5mm,width=4.5mm]}]
    (12.52,-4.95) -- (13.98,-4.95);

\node[font=\bfseries\fontsize{22}{22}\selectfont,text=mainblue]
    at (11.70,-4.95) {$\cdots$};

\draw[eventred,line width=4.5pt] (3.10,-4.95) -- (4.25,-4.95);
\draw[eventred,line width=4.5pt] (6.00,-4.95) -- (7.5,-4.95);

\draw[eventgold,line width=4.5pt] (6.82,-4.95) -- (9.35,-4.95);
\draw[eventgold,line width=4.5pt,opacity=0.7] (6.3,-4.95) -- (6.82,-4.95);
\draw[eventred,decorate,decoration={brace,amplitude=5pt},line width=.9pt]
    (3.10,-4.65) -- (4.25,-4.65)
    node[midway,above=8pt,eventfont] {Event 1};

\draw[eventred,decorate,decoration={brace,amplitude=5pt},line width=.9pt]
    (6.00,-4.65) -- (6.88,-4.65)
    node[midway,above=8pt,eventfont] {Event 2};

\draw[eventred,decorate,decoration={brace,amplitude=5pt},line width=.9pt]
    (6.3,-4.85) -- (9.35,-4.85)
    node[pos=.55,above=8pt,eventfont] {Event 3};

\draw[line width=.5pt] (3.08,-4.95) -- (3.08,-5.28);
\draw[line width=.5pt] (4.28,-4.95) -- (4.28,-5.28);
\draw[line width=.5pt] (5.98,-4.95) -- (5.98,-5.28);
\draw[line width=.5pt] (6.82,-4.95) -- (6.82,-5.28);
\draw[line width=.5pt] (9.38,-4.95) -- (9.38,-5.28);

\node[timelab,anchor=north] at (3.10,-5.28) {$x_1$};
\node[timelab,anchor=north] at (4.25,-5.2) {$x_1+d_1$};

\node[timelab,anchor=north] at (6.00,-5.28) {$x_2$};
\node[timelab,anchor=north] at (7.50,-5.2) {$x_2+d_2$};
\draw[line width=.5pt] (6.3,-4.95) -- (6.3,-5.28);
\node[timelab,anchor=north] at (6.5,-5.3) {$x_3$};
\node[timelab,anchor=north] at (9.35,-5.2) {$x_3+d_3$};

\draw[epochorange,line width=1.2mm] (1.38,-6.45) -- (10.90,-6.45);
\draw[epochorange,line width=1.2mm,-{Triangle[length=4mm,width=4mm]}]
    (12.52,-6.45) -- (13.98,-6.45);


\node[font=\bfseries\fontsize{22}{22}\selectfont,text=mainblue]
    at (11.75,-6.45) {$\cdots$};

\foreach \xx in {3.45,4.95,6.40,7.80,9.20,10.55}{
    \draw[line width=.5pt] (\xx,-6.26) -- (\xx,-6.64);
}

\draw[mainblue,line width=.5pt] (4.95,-6.26) -- (4.95,-6.64);

\node[font=\itshape\fontsize{10}{10}\selectfont]
    at (5.65,-6.00) {30 seconds};

\foreach \a/\b/\lab in {
    1.72/3.45/1,
    3.45/4.95/2,
    4.95/6.40/3,
    6.40/7.80/4,
    7.80/9.20/5,
    9.20/10.55/6
}{
    \draw[
        orange!85!black,
        decorate,
        decoration={brace,amplitude=5pt,mirror},
        line width=.75pt
    ] (\a,-6.55) -- (\b,-6.55)
      node[midway,below=10pt,epochfont] {Epoch \lab};
}

\node[anchor=west,font=\fontsize{13}{13}\selectfont]
    at (12.72,-7.08) {Time};

\end{tikzpicture}

}
\caption{From PSG to BEST and QEL semantics. (A) Exemplar PSG segment. (B) PSG temporal structure and QEL metric operators ($\Diamond_t,\Box_t$) evaluated on a dense timeline, inducing a BEST represented by labeled intervals $[x_i,x_i+d_i)$ alongside standard 30\,s epoch staging.}
\label{fig:psg}
\vspace{0.3\baselineskip}
\end{figure}

The scoring of these recordings is governed by the AASM manual, which defines technical specifications, signal patterns and temporal rules for interpretation. The semantic annotation of PSG necessitates the integration of outcomes across two distinct temporal granularities. The first category comprises discrete clinical events (e.g., apneas, periodic limb movements, arousals), which are defined over variable-length intervals measured in seconds. The second category is sleep stages, which is analyzed based on fixed-length 30-second epochs. The scoring spans the Total Recording Time (TRT), delimited by the 'Lights Out' (start) and 'Lights On' (end) markers. 

Despite shared criteria, substantial inter-scorer variability persists. Studies report overall stage agreement of $\sim$80--85\% \cite{inter_program}. However, agreement is lower for specific stages, with $\sim$63\% for N1 and $\sim$67\% for N3 \cite{inter_program}. Agreement is also limited for event scoring. Expert-to-expert agreement ($\kappa$) for arousal detection ranges from 0.63 to 0.67, and agreement for respiratory events ranges from 0.76 to 0.79 \cite{caisr}. A key driver is that many directives remain qualitative (e.g., ``majority of the epoch,'' ``associated with'') rather than formally defined logic. Such wording leaves room for subjective judgment and inconsistent outcomes across scorers. This ambiguity also becomes a computational bottleneck. Sleep analysis tools and cohort discovery pipelines often resolve underspecified directives using hard-coded heuristics that reflect individuals' interpretations, which reduces reproducibility across institutions and studies. 

\subsection*{2.2\quad Preliminaries: Ensemble Logic over the Rational Numbers (QEL)}

We use QEL to represent sleep-scoring rules with explicit timing constraints. QEL is a temporal logic defined over a dense timeline, where time points range over $\mathbb{Q}_{\ge 0}$. This choice is natural for PSG because physiological signals are continuously recorded and annotated with rational-valued onset and offset times.

\begin{wraptable}{l}{0.56\textwidth}
\vspace{-0.3\baselineskip}
\captionsetup{font=small,skip=1pt}
\centering
\footnotesize
\setlength{\tabcolsep}{3pt}
\renewcommand{\arraystretch}{1.03}

\caption{Correspondence between sleep-study entities and QEL/BEST}
\label{tab:correspondence}

\begin{tabular}{@{}>{\raggedright\arraybackslash}p{0.19\textwidth}
                >{\raggedright\arraybackslash}p{0.14\textwidth}
                >{\raggedright\arraybackslash}p{0.19\textwidth}@{}}
\toprule
\textbf{Sleep-study entities} & \textbf{Logical constructs} & \textbf{Examples} \\
\midrule

AASM terms \& PSG-derived features 
& Atomic propositions ($\sf P$) 
& $\event{Flowdrop90}$, $\event{Desat3}$, $\event{EEGFreqGT16}$, $\event{Alpha}$, $\event{Spindle}$ \\
\midrule

PSG records 
& Semantic model (BEST) 
& $\varepsilon:\sf P\to {\sf IE}$ \\
\midrule

Time-indexed labeling 
& Interpretation induced by BEST 
& $\alpha_\varepsilon:\mathbb{Q}\to 2^{\sf P}$, 
$\alpha_\varepsilon(x)=\{p\in\sf P\mid x\in \bigcup \varepsilon(p)\}$ \\
\midrule

Temporal relationships 
& Modal and logical operators 
& $\varphi_t$, $\Box_t\varphi$, $\Diamond_t\varphi$, $\wedge$, $\vee$, $\neg$, $\exists x$, $\forall x$ \\
\midrule

Sleep events library 
& QEL formulas 
& $\event{Arousal}$, $\event{CentralApnea}$ \\
\midrule

Sleep patterns 
& QEL formulas 
& $\event{SleepTrans}$, $\event{SleepFrag}$ \\
\midrule

Execution 
& Model checking 
& $(\varepsilon,q)\models\varphi$ \\

\bottomrule
\end{tabular}
\vspace{0pt} 
    \parbox{0.56\textwidth}{
        \footnotemark \scriptsize QEL, Rational Ensemble Logic; BEST, Biomedical Event Structure Temporal Model; IE, interval ensemble. Examples are illustrative rather than exhaustive. Atomic propositions correspond to signal-derived observations; higher-level sleep events and sleep patterns are represented as QEL formulas. 
    }
    \vspace{-15pt}
\end{wraptable}
A QEL formula is built from Atomic propositions, Boolean connectives, first-order quantifiers, and metric temporal operators. Atomic propositions represent clinically irreducible observations or signal-derived events, such as spindles and oxygen desaturations, which are treated as the smallest meaningful units for formalization. Boolean operators ($\wedge,\vee,\neg$) express conjunction, disjunction, and negation. First-order quantifiers ($\exists x,\forall x$) allow formulas to refer to time anchors and durations explicitly over time. 

QEL uses three key temporal operators: (i) displacement $\varphi_t$ shifts the evaluation of a formula by a temporal offset $t$, representing $\varphi$ is true t-units of time away.  (ii) metricized modal operator $\Box_t \varphi$ states that $\varphi$ holds always within duration $t$. (iii) $\Diamond_t \varphi$ states that $\varphi$ happens sometime within duration $t$. Together, these constructs allow QEL to express the timing patterns that frequently appear in AASM rules, such as ``lasting at least 10 seconds,'' ``preceded by,'' or ``occurs within an epoch.''

In this paper, QEL is used as the symbolic language for encoding sleep-scoring rules, while the underlying PSG record is represented separately as a BEST.

\subsection*{2.3\quad Biomedical Event Structure Temporal Model (BEST)}

We represent a scored PSG record as a Biomedical Event Structure Temporal Model (BEST). Formally,  we define an Interval Ensemble (IE) as a finite set of non-overlapping intervals on the timeline (e.g., all intervals where a PSG recording was annotated as  apnea). A BEST ($\varepsilon$) is a mapping from Atomic Propositions to interval ensembles.

Formally, each proposition in a proposition set $\sf P$ is associated with a finite set of non-overlapping intervals over $\mathbb{Q}_{\ge 0}$. Thus, a BEST can be viewed as a mapping $\varepsilon : \sf P \to {\sf IE}$. For each proposition $p\in\sf P$, $\varepsilon(p)$ specifies the intervals during which $p$ holds.

This representation is well aligned with PSG annotation practice, since sleep events are naturally stored as labeled intervals rather than isolated time points. A BEST therefore serves as the semantic model on which QEL formulas are evaluated. Given a BEST $\varepsilon$ and a reference time $t$, model checking determines whether a QEL formula $\varphi$ holds at that anchor, written as $(\varepsilon,t)\models\varphi$.

In summary, QEL provides the formal language for expressing sleep-scoring rules, and BEST provides the interval-based representation of PSG records against which those rules are evaluated.

\subsection*{2.4 Representative AASM rules used in this study}\label{sec:rule_contexts}
This work focuses on translating representative AASM scoring directives into executable QEL specifications. Below we summarize the clinical intent and key temporal constraints for the related domains used throughout the paper based on AASM manual (version 2.4, 2017).

\begin{wrapfigure}{l}{0.6\textwidth}
    \centering
    \vspace{-10pt}
    \resizebox{\linewidth}{!}{
\begin{tikzpicture}[>=stealth, font=\sffamily, thick]
    \node[draw=black!80, fill=orange!10, text width=11cm, inner sep=10pt, rounded corners=2pt, align=left] (rule) at (0,0) {
        \centering \textbf{Scoring Rules}\par\vspace{4pt}
        Score event 1 if \textcolor{blue!70!black}{\textbf{somewhere}} \colorbox{green!70!black}{\textcolor{white}{\textbf{observation 1}}} appears \textcolor{blue!70!black}{\textbf{followed by at least 3 second}} of \colorbox{green!70!black}{\textcolor{white}{\textbf{observation 2}}} \textcolor{orange!90!black}{\textbf{and}} \colorbox{green!70!black}{\textcolor{white}{\textbf{observation 3}}} \textcolor{orange!90!black}{\textbf{not}} happen \textcolor{blue!70!black}{\textbf{throughout}} the events
    };

    \node[draw=green!50!black, fill=white!90, text width=3.2cm, align=center, inner sep=0pt, minimum height=2.5cm] (c1) at (-3.8, -3) {};
    \node[fill=green!50!black, text=white, minimum width=3.2cm, minimum height=0.8cm, font=\bfseries, align=center] at ([yshift=-0.4cm]c1.north) {Atomic/derived\\Propositions};
    \node[text=green!50!black, align=left] at ([yshift=-1.6cm]c1.north) {Observation 1: $p_1$\\Observation 2: $p_2$\\Observation 3: $p_3$};

    \node[draw=blue!70!black, fill=white!90, text width=3.2cm, align=center, inner sep=0pt, minimum height=2.5cm] (c2) at (0, -3) {};
    \node[fill=blue!70!black, text=white, minimum width=3.2cm, minimum height=0.8cm, font=\bfseries, align=center] at ([yshift=-0.4cm]c2.north) {Temporal\\Relationships};
    \node[text=blue!50, align=center, font=\small] at ([yshift=-1.6cm]c2.north) {Somewhere $\Diamond$\\[4pt] throughout $\Box$\\[4pt] 3 second: time terms};

    \node[draw=orange!90!black, fill=white!90, text width=3.2cm, align=center, inner sep=0pt, minimum height=2.5cm] (c3) at (3.8, -3) {};
    \node[fill=orange!90!black, text=white, minimum width=3.2cm, minimum height=0.8cm, font=\bfseries, align=center] at ([yshift=-0.4cm]c3.north) {Logic Operators};
    \node[text=orange!80, align=center] at ([yshift=-1.6cm]c3.north) {NOT $\neg$\\[4pt] AND $\wedge$};

    \draw[->, line width=2pt, green!50!black] (-3.8, -1.2) -- (-3.8, -1.75);
    \draw[->, line width=2pt, blue!70!black] (0, -1.2) -- (0, -1.75);
    \draw[->, line width=2pt, orange!90!black] (3.8, -1.2) -- (3.8, -1.75);

    \node[draw=blue!30!black, fill=white!90, text=white, minimum width=10.2cm, minimum height=1.5cm, align=center, outer sep=0] (formula) at (0, -5.5) {};
    \node[fill=white, inner sep=4pt] at (0, -5.3) {
        \textcolor{orange!90!black}{$\exists x \exists u \exists t$} \Big(
        (\textcolor{blue!70!black}{$\Diamond_t$} \textcolor{green!60!black}{$p_1$})$_x$ \textcolor{orange!90!black}{$\wedge$}
        (\textcolor{blue!70!black}{$\Box_u$} \textcolor{green!60!black}{$p_2$})$_{x+t}$ \textcolor{orange!90!black}{$\wedge$}
        (\textcolor{blue!70!black}{$\Box_{t+u}$} \textcolor{orange!90!black}{$\neg$} \textcolor{green!60!black}{$p_3$})$_x$ \Big)
    };
    \node[fill=white, inner sep=2pt] at (0, -5.9) {with \textcolor{blue!70!black}{$u \ge 3, x, u, t \in \mathbb{Q}_{\ge 0}$}};

    \draw[->, line width=2pt, green!50!black] (-3.8, -4.25) -- (-3.8, -4.75);
    \draw[->, line width=2pt, blue!70!black] (0, -4.25) -- (0, -4.75);
    \draw[->, line width=2pt, orange!90!black] (3.8, -4.25) -- (3.8, -4.75);

\draw[->, dashed, line width=1.5pt, blue!70!black] (formula.west) -- ++(-2.2,0) |- ([yshift=-0.2cm]rule.west);

\node[fill=white, draw=blue!70!black, thick, rounded corners=3pt, text=blue!70!black, font=\bfseries] at (-7.3, -2.8) {Translation};
\end{tikzpicture}
}
    \caption{Overview of the AASM-to-QEL translation workflow.}
    \label{fig:aasm_tel_workflow}
    \vspace{0.3\baselineskip}
\end{wrapfigure}
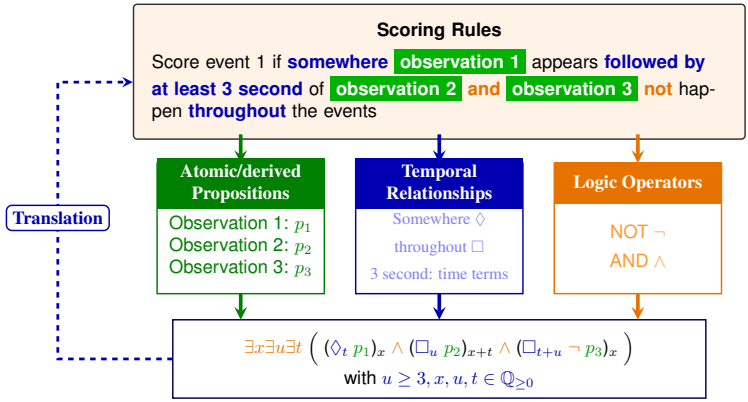
\par\addvspace{0.25em}

\textbf{Arousal (AASM V.A.1).}: An arousal is scored during sleep when there is an abrupt EEG frequency shift (alpha, theta, or $>16$\,Hz, excluding spindles) lasting at least 3\,s, preceded by at least 10\,s of stable sleep; during REM, a concurrent increase in submental EMG lasting at least 1\,s is additionally required.

\textbf{Apnea and Hypopnea  (AASM VIII.C--D).}: Apnea is defined by a $\ge 90\%$ airflow reduction for $\ge 10$\,s with subtype classification by inspiratory effort (present throughout: obstructive; absent throughout: central; absent initially then present: mixed), and events entirely during wake are not scored. Hypopnea is defined by a $\ge 30\%$ airflow reduction for $\ge 10$\,s plus either desaturation or arousal; in practice, desaturation/arousal may lag the airflow reduction due to physiologic latency, motivating an explicit delay window in formalization.

\textbf{Stage N2 (AASM IV.G).}: N2 scoring is an epoch-based process initiated by the appearance of sleep spindles or K-complexes. Once initiated, this state is maintained across subsequent epochs characterized by low-amplitude, mixed-frequency (LAMF) EEG activity. The N2 state explicitly terminates upon transitioning to stage W, N3, or REM sleep, or interrupted by an arousal or a major body movement that is not followed by characteristic N2 waveforms. Scoring edge cases involve: (i) N2 markers are followed by an arousal within the same  subsequent epoch; and (ii) macroscopic evaluations requiring the N2 state to account for the majority of the current epoch. AASM Manual (version 2.4, Section IV.G, Figure 7, p.23) summarizes canonical AASM staging scenarios that motivate the edge cases addressed by the N2 scoring formulation.

\textbf{Limb movements and PLM series (AASM VII.B).}: A leg movement (LM) is defined by an anterior tibialis EMG increase above baseline with duration constraints (0.5--10\,s) and an explicit termination criterion (return below threshold for at least 0.5\,s). LMs are excluded if they occur during respiratory events or within a short temporal proximity window. A periodic limb movement (PLM) series is defined as at least four qualifying LMs with inter-movement onset intervals (period lengths) between 5 and 90\,s.

\section*{3. Method}\label{sec:methods}

\subsection*{3.1 \quad Conceptual correspondence from PSG to QEL}

We model time as elapsed seconds from recording start over the nonnegative rationals $\mathbb{Q}_{\ge 0}$. A scored PSG record is represented as a BEST, $\varepsilon:\sf P \to {\sf IE}$, where $\sf P$ is a finite set of atomic propositions and ${\sf IE}$ denotes interval ensembles. For each $p\in\sf P$, $\varepsilon(p)$ specifies the intervals during which $p$ holds, each represented by its onset and offset times. For example, $\event{Spindle}$ is true exactly on intervals where spindle activity is present, while propositions absent from a record map to the empty ensemble. A BEST induces the usual time-indexed interpretation $\alpha_\varepsilon(x)=\{\,p\in\sf P \mid x\in \bigcup \varepsilon(p)\,\}$, which records the propositions true at time $x$. In this work, however, BEST is the primary representation because PSG annotations are naturally interval-based.

As shown in Fig.\ref{fig:psg} (B), a scored PSG session is finite and bounded by clinically defined $\event{LightOut}$ and $\event{LightOn}$ timestamps, denoted $T_{\texttt{start}}$ and $T_{\texttt{end}}$, with $0\le T_{\texttt{start}}< T_{\texttt{end}}$. Accordingly, QEL formulas are evaluated over the bounded session interval $[T_{\texttt{start}},T_{\texttt{end}}]$, with temporal windows interpreted relative to that extent. Additional domain constraints, such as mutual exclusivity of sleep stages and hierarchical dependencies among respiratory labels, are enforced separately as global invariants. 

To rigorously encapsulate the AASM directive that clinical sleep staging must be executed in "30-second, sequential epochs commencing at the start of the study", we define an ordered set of epoch onset times $E=\{e_0,e_1,\ldots,e_n\} $ (where $e_i = e_0 + iL$ and $L=30$ second is the epoch length) for any PSG records. The epoch-level scoring requirement imposes a strict monochromaticity constraint on the semantic model for any staging proposition $\sf S \in \{Wake, N1,N2,N3,REM, Movements\}$. Specifically, for any epoch interval $I := [a,b)$ with a fixed temporal duration $L = 30\,\mathrm{s}$, if the epoch onset point $a$ is labeled with a specific sleep stage $\sf S$, the proposition $S$ must logically evaluate to true at every continuous time point $x \in I$ (i.e., $\sf \sf S_a \rightarrow \Box_L \sf S_a)$. Crucially, staging is mutually exclusive: for all $x\in[e_i,e_{i+1})$, exactly one stage label is active.

Table~\ref{tab:correspondence} summarizes the correspondence between sleep study entities and their QEL/BEST representations. In particular, signal-derived events are encoded as atomic propositions, AASM-defined events are formalized as QEL formulas, and event recognition is performed by model checking, i.e., determining whether $(\varepsilon,q)\models\varphi$ at an anchor time $q$.

\subsection*{3.2\quad Formalizing AASM rules with QEL}

We translate AASM scoring rules (version 2.4, 2017) \cite{berry2017aasm, berry2017aasm_update} expressed in natural language into executable QEL specifications using a structured extraction-and-compilation procedure (Fig.~\ref{fig:aasm_tel_workflow}). Each translated event $E$ is represented by a QEL formula $\varphi_E$ that holds exactly at those anchor times where $E$ occurs in a BEST-encoded record. The procedure consists of three steps.

\textbf{Step 1: Atomic proposition extraction.} For each target rule, we extract a finite set of auditable observations referenced by the rule text and PSG signals, using them as atomic propositions in $\sf P$. Threshold-based features (e.g., amplitude, frequency) are direct APs.  Clinically defined waveform-morphology patterns (e.g., spindles, alpha rhythm) are also treated as APs rather than further decomposed, since finer primitives often lack clear clinical meaning.  When the AASM rule references higher-level clinical context (e.g., staging propositions, $\event{Sleep}$, $\event{NREM}$, $\event{REM}$), we introduce \emph{derived propositions} as named QEL macros built from APs and previously constructed event formulas.
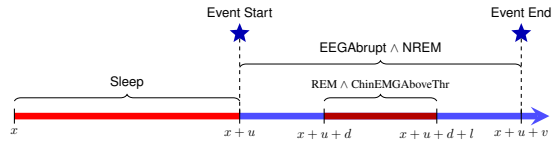
\begin{wrapfigure}{l}{0.45\textwidth}
    \centering
\resizebox{0.99\linewidth}{!}{
\begin{tikzpicture}[>=stealth, font=\sffamily, x=1.5cm, y=1cm]
    \draw[line width=5pt, red] (0,0) -- (4,0);
    \draw[line width=5pt, blue!70!white] (4,0) -- (9,0);
    \draw[line width=5pt, red!70!black] (5.5,0) -- (7.5,0);
    \draw[thick, ->, blue!70!white, line width=5pt] (9,0) -- (9.5,0);

    \draw[thick] (0, 0.2) -- (0, -0.2) node[below] {$x$};
    \draw[thick] (4, 0.2) -- (4, -0.2) node[below] {$x+u$};
    \draw[thick] (5.5, 0.2) -- (5.5, -0.2) node[below, align=center] {$x+u+d$};
    \draw[thick] (7.5, 0.2) -- (7.5, -0.2) node[below, align=center] {$x+u+d+l$};
    \draw[thick] (9, 0.2) -- (9, -0.2) node[below] {$x+u+v$};

    \draw[decorate, decoration={brace, amplitude=5pt}] (0, 0.4) -- (4, 0.4) node[midway, above=6pt] {Sleep};
    \draw[decorate, decoration={brace, amplitude=5pt}] (4, 1.3) -- (9, 1.3) node[midway, above=6pt] {EEGAbrupt $\wedge$ NREM};
    \draw[decorate, decoration={brace, amplitude=5pt}] (5.5, 0.4) -- (7.5, 0.4) node[midway, above=6pt, font=\small] {REM $\wedge$ ChinEMGAboveThr};

    \node[star, star points=5, star point ratio=2.25, fill=blue!70!black, inner sep=2.5pt, label=above:{Event Start}] (start) at (4, 2.2) {};
    \node[star, star points=5, star point ratio=2.25, fill=blue!70!black, inner sep=2.5pt, label=above:{Event End}] (end) at (9, 2.2) {};
    \draw[dashed, thick] (start) -- (4, 0);
    \draw[dashed, thick] (end) -- (9, 0);
\end{tikzpicture}
}
\caption{Semantic topology of the AASM Arousal scoring rule formalized in QEL.}
\label{fig:arousal_TEL}
\end{wrapfigure}

\textbf{Step 2: Temporal structure extraction.}
We parse the rule narrative into an explicit temporal dependency graph at the time anchor where the rule is to be evaluated. In the AASM rules, this structure typically consists of (i) sequential ordering (e.g., ``preceded by'' or ``following'' ), which establishes geometric distance between phenotypes along the timeline;  (ii) continuous persistence, which is a local evidence window in which a phenotype must persist for a bounded duration (e.g., ``lasting at least $t$ seconds'',``throughout'' or ``continuous''); (iii) bounded existential occurrence (e.g., ``there is ...'' or ``associated with''), which capture transient micro-events or physiological responses within a bounded window. We make this structure explicit as a temporal dependency graph and then map each clause into QEL operators using the correspondence in Table~\ref{tab:correspondence}. Specifically, we use $\Box_t$ for bounded persistence , $\Diamond_t$ for bounded existence, and  $\varphi_u$ for displacement.  Fig. \ref{fig:arousal_TEL} illustrates a temporal structure of arousal rules defined in AASM Manual.

\textbf{Step 3: QEL formula construction.}
We construct the final QEL formula by combining the extracted APs and temporal structures. First, we introduce time-length variables to parameterize temporal distances, subjecting these quantified variables to explicit algebraic side constraints to rigorously enforce the thresholds defined in the clinical rules. Second, concurrency and condition branching are resolved via Boolean operators.  Lastly, we translate ambiguous clinical descriptions into precise mathematical logic. For example, we represent ``associated with'' by ($\Diamond_t$) to indicate that an event occurs within a bounded time window.

\subsection*{3.3 \quad LLM-Assisted Back-Translation Evaluation}
\begin{wraptable}{l}{0.5\textwidth}
\centering
\footnotesize
\setlength{\tabcolsep}{4pt}
\renewcommand{\arraystretch}{1.06}
\caption{Atomic propositions extracted in this study.}
\label{tab:propositions}
\begin{tabular}{@{}ll@{}}
\toprule
\textbf{Proposition} & \textbf{Clinical meaning / role} \\
\midrule

$\event{Alpha}$ & Alpha EEG activity is present \\
$\event{Theta}$ & Theta EEG activity is present \\
$\event{EEGFreqGT16}$ & EEG frequency greater than 16 Hz is present \\
$\event{Spindle}$ & Sleep spindle is present \\
$\event{KC}$ & K-complex is present \\
$\event{ChinEMGAboveThr}$ & Chin/submental EMG exceeds threshold \\
$\event{MajorBodyMove}$ & Major body movement is present \\

$\event{FlowDrop90}$ & \begin{tabular}{@{}l@{}}Airflow / signal excursion reduction \\ meeting the 90\% threshold\end{tabular} \\
$\event{FlowDrop30}$ & \begin{tabular}{@{}l@{}}Airflow / signal excursion reduction \\ meeting the 30\% threshold\end{tabular} \\
$\event{FlowBelowThr}$ & Inspiratory flow flattening / reduced flow is present \\
$\event{EffortAboveThr}$ & Inspiratory effort is present above threshold \\
$\event{Snore}$ & Snoring signal is present \\
$\event{PhaseDiffAboveThr}$ & \begin{tabular}{@{}l@{}}Thoracoabdominal paradox  \\ phase difference is present\end{tabular} \\
$\event{Desat3}$ & Oxygen desaturation meeting the 3\% threshold \\
$\event{Desat4}$ & Oxygen desaturation meeting the 4\% threshold \\

$\event{EMGAboveRestThr}$ & Leg EMG is above resting threshold \\
$\event{EMGAbove8}$ & Leg EMG is at least $8\,\mu V$ above resting EMG \\
$\event{EMGAbove2}$ & Leg EMG exceeds $2\,\mu V$ above resting EMG \\

\bottomrule
\end{tabular}
\end{wraptable}

To assess whether QEL specifications preserve the semantic content of the original narrative AASM rules, we conducted a back-translation experiment using the pinned OpenAI snapshot model \textit{gpt-5-mini-2025-08-07} via the Responses API with its default settings. For each selected rule, the model was provided with the corresponding QEL formula, a controlled glossary of atomic propositions, and, when needed, structured macro definitions for referenced derived predicates. The model was instructed to generate a clinician-facing natural-language description that preserved thresholds, durations, negations, and temporal dependencies; a dummy example was included only to specify the expected output format. To minimize prompt leakage, the model was not shown the source AASM text during generation, was not given access to external references, and did not receive prompts containing the keyword ``AASM.'' Unless otherwise specified, all generation parameters were left at their default values.

Semantic fidelity was quantified using OpenAI \textit{text-embedding-3-small} (1536-dimensional output). Both the generated back-translations and the corresponding source narratives were embedded using default settings, without explicitly setting the output dimensionality and using the default floating-point encoding. Semantic similarity was defined as cosine similarity between the generated and reference embeddings. Lexical overlap was evaluated separately using sequence matching ratio and ROUGE-L.

\section*{4. Results} 

\subsection*{4.1 \quad Formalization of  Sleep Events}

We instantiated the proposed AASM-to-QEL translation framework across four representative scoring domains: arousal, respiratory events, stage N2, and limb movements/PLMS. Following the extraction-and-compilation procedure, we extracted 18 unique atomic propositions (Table~\ref{tab:propositions}). Using these atomic propositions, we derived 18 QEL rule specifications, including 12 final scoring specifications corresponding to clinically scoreable AASM events. Table~\ref{tab:unified_sleep_rules} summarizes these translation results.

These specifications are not literal, term-by-term transcriptions of the AASM Manual. Instead, they formalize the parsed temporal and logical topology of the scoring rules. Specification rule (6)  in Table~\ref{tab:unified_sleep_rules} illustrates why direct textual translation can be inadequate. The AASM Manual states that the event is associated with an arousal or desaturation, but a literal interpretation as immediate temporal coincidence may not capture the physiologic delay implied in clinical practice. We therefore introduce a delay term, $\varepsilon$, to model this latency and better align the formal specification with the intended clinical meaning.

\par\addvspace{1em}
{\footnotesize
\setlength{\jot}{0pt}
\setlength{\tabcolsep}{3pt}
\renewcommand{\arraystretch}{1.35}
\setlength{\LTpre}{12pt}
\setlength{\LTpost}{12pt}
\begin{longtable}{@{}>{\raggedright\arraybackslash}p{0.38\textwidth}
                  >{\raggedright\arraybackslash}p{0.14\textwidth}
                  >{\raggedright\arraybackslash}p{0.42\textwidth}@{}}
\caption{QEL formalization of representative AASM rules for arousal, apnea, hypopnea, stage N2, leg movement (LM), and periodic limb movement series (PLMS). Rule 1-9, 15, and 17-18 are final scoring rules. $E$ denotes the epoch onset time set, $L=30$ seconds denotes epoch length. $u$,$v$,$\ell$ and $d$ are time-length terms and $x$ is time variables}
\label{tab:unified_sleep_rules}\\

\toprule
\textbf{Rules} & \textbf{Atomic/derived propositions} & \textbf{Formula} \\
\midrule
\endfirsthead

\toprule
\textbf{Rules} & \textbf{Atomic/derived propositions} & \textbf{Formula} \\
\midrule
\endhead

\midrule
\multicolumn{3}{r}{\textit{Continued on next page}} \\
\endfoot

\bottomrule
\endlastfoot

\multicolumn{3}{@{}l}{\textbf{Arousal (AASM V.A.1)}} \\

\textbf{(1) Arousal:} during sleep, an abrupt EEG frequency shift (alpha, theta, or $>16$\,Hz, excluding spindles) lasting at least 3\,s, preceded by at least 10\,s of stable sleep; if in REM, concurrent chin EMG increase lasting at least 1\,s is additionally required. &
$\event{Alpha}$, $\event{Theta}$, $\event{EEGFreqGT16}$, $\event{Spindle}$, $\event{ChinEMGAboveThr}$ &
$\displaystyle
\varphi_{\event{Arousal}} := \exists x\,\exists u\,\exists v\,\exists \ell \Big(
(\Box_u\,\event{Sleep})_{x}
\wedge
(\Box_v\,((\event{Alpha} \vee \event{Theta} \vee \event{EEGFreqGT16}) \wedge \neg \event{Spindle}))_{x+u}
\wedge
\big(
\event{NREM}_{x+u}
\vee
(\event{REM}_{x+u}\wedge \Diamond_v(\Box_{\ell}\,\event{ChinEMGAboveThr})_{x+u})
\big)
\Big),
\; u\ge10,\; v\ge3,\; \ell\ge1
$ \\
\hline
\multicolumn{3}{@{}l}{\textbf{Apnea (AASM VIII.C)}} \\

\textbf{(2) Apnea:} airflow drop $\ge 90\%$ from baseline for $\ge 10$\,s; events occurring entirely during wake are not scored. &
$\event{FlowDrop90}$, $\event{Wake}$ &
$\begin{aligned}[t]
&\varphi_{\text{apnea}}:= \exists x \exists u \Big( (\Box_u \event{FlowDrop90})_x \wedge \neg(\Box_u \event{Wake})_x \Big),\\
&u \ge 10
\end{aligned}$ \\

\textbf{(3) Obstructive apnea:} meets apnea criteria and inspiratory effort is present throughout the event. &
$\event{FlowDrop90}$, $\event{Wake}$, $\event{EffortAboveThr}$ &

$\begin{aligned}[t]
&\varphi_{\text{obs-apnea}} := \exists x \exists u \Big(
(\Box_u \event{FlowDrop90})_x
\wedge
\neg(\Box_u \event{Wake})_x \\
&\wedge
(\Box_u \event{EffortAboveThr})_x
\Big),
\; u \ge 10  
\end{aligned}
$ \\

\textbf{(4) Central apnea:} meets apnea criteria and inspiratory effort is absent throughout the event. &
$\event{FlowDrop90}$, $\event{Wake}$, $\event{EffortAboveThr}$ &
$\begin{aligned}[t]
&\varphi_{\text{cen-apnea}} := \exists x \exists u \Big(
(\Box_u \event{FlowDrop90})_x
\wedge
\neg(\Box_u \event{Wake})_x \\
&\wedge
(\Box_u \neg \event{EffortAboveThr})_x
\Big),
\; u \ge 10
\end{aligned}
$ \\

\textbf{(5) Mixed apnea:} meets apnea criteria; effort is absent initially and then resumes later in the same apnea. &
$\event{FlowDrop90}$, $\event{Wake}$, $\event{EffortAboveThr}$ &
$\displaystyle
\varphi_{\text{mixed-apnea}} := \exists x \exists u \exists v \exists \ell \Big(
(\Box_u \event{FlowDrop90})_x
\wedge
\neg(\Box_u \event{Wake})_x
\wedge
(\Box_v \neg \event{EffortAboveThr})_x
\wedge
(\Box_{\ell} \event{EffortAboveThr})_{x+v}
\Big),
\; u \ge 10,\; v+\ell=u
$ \\
\hline
\multicolumn{3}{@{}l}{\textbf{Hypopnea (AASM VIII.D)}} \\

\textbf{(6) Hypopnea (1A, recommended):} airflow reduction $\ge 30\%$ for $\ge 10$\,s plus either $\ge 3\%$ desaturation or arousal, allowing physiologic latency. &
$\event{FlowDrop30}$, $\event{Desat3}$ &
$
\varphi_{\text{hyp-1A}} := \exists x \exists u \Big(
(\Box_u \event{FlowDrop30})_x
\wedge
\big(
(\Diamond_{u+\epsilon}\event{Desat3})_x
\vee
(\Diamond_{u+\epsilon}\varphi_{\event{Arousal}})_x
\big)
\Big),
\; u \ge 10
$ \\

\textbf{(7) Hypopnea (1B, acceptable):} airflow reduction $\ge 30\%$ for $\ge 10$\,s plus $\ge 4\%$ desaturation. &
$\event{FlowDrop30}$, $\event{Desat4}$ &
$\begin{aligned}[t]
&\varphi_{\text{hyp-1B}}:= \exists x \exists u \Big( (\Box_u \event{FlowDrop30})_x \wedge (\Diamond_{u+\epsilon}\event{Desat4})_x \Big), \\
& u \ge 10
\end{aligned}$ \\

\textbf{(8) Obstructive hypopnea:} qualifying hypopnea with snoring, inspiratory flow flattening, or thoracoabdominal paradox. &
$\event{Snore}$, $\event{FlowBelowThr}$, $\event{PhaseDiffAboveThr}$ &
$\begin{aligned}[t]
&\varphi_{\text{obs-hyp}}:= \varphi_{\text{hyp-1A/hyp-1B}} \wedge \exists x \exists v\Big((\Diamond_v(\event{Snore} \\
&\vee \event{FlowBelowThr} \vee \event{PhaseDiffAboveThr}))_x \Big),\quad 0 \le v \le u 
\end{aligned}$ \\

\textbf{(9) Central hypopnea :} qualifying hypopnea with none of the obstructive features present during the scored window. &
$\event{Snore}$, $\event{FlowBelowThr}$, $\event{PhaseDiffAboveThr}$ &
$\begin{aligned}[t]
&\varphi_{\text{cen-hyp}} := \varphi_{\text{hyp-1A/hyp-1B}} \wedge \exists v \exists x\Big( (\Box_v \neg(\event{Snore} \\
&\vee \event{FlowBelowThr}\vee \event{PhaseDiffAboveThr}))_x \Big), \quad 0 \le v \le u
\end{aligned}$ \\
\hline
\multicolumn{3}{@{}l}{\textbf{Stage N2 (AASM IV.G)}} \\

\textbf{(10) Valid K-complex:} K-complex used for N2 staging must be unassociated with arousal. &
$\event{KC}$ &
$\begin{aligned}[t]
&\varphi_{\text{ValidKC}}:= \exists d \exists x\Big( (\Box_d \event{KC})_x \wedge (\neg \event{KC})_{x+d} \\ 
&\wedge (\Box_{d+1}\neg \varphi_{\event{Arousal}})_x \Big), \quad d \ge 0.5
\end{aligned}$ \\

\textbf{(11) N2 signal:} characteristic N2 waveform used by the downstream rules. &
$\event{Spindle}$, $\event{KC}$ &
$\displaystyle
\varphi_{\text{Sig}} := \varphi_{\text{ValidKC}} \vee \event{Spindle}
$ \\

\textbf{(12) N2 initiation from characteristic waveforms:} N2 is initiated if a characteristic N2 signal occurs in the first half of the current epoch or the second half of the previous epoch, with no arousal in the current epoch. &
$\event{Spindle}$, $\event{KC}$ &
$\displaystyle
\varphi_{\text{N2InitSig}} := \exists x \Big(
\big(
(\Diamond_{L/2}\varphi_{\text{Sig}})_x
\vee
(\Diamond_{L/2}\varphi_{\text{Sig}})_{x-L/2}
\big)
\wedge
(\Box_L \neg \varphi_{\event{Arousal}})_x
\Big)
$ \\

\textbf{(13) N2 initiation from N3:} after an epoch scored as N3, the current epoch is scored as N2 when N3 no longer persists and no higher-priority interruption occurs. &
$\event{N3}$, $\event{Wake}$, $\event{REM}$ &
$\displaystyle
\varphi_{\text{N2InitFromN3}} := \exists x\Big(
(\event{N3})_{x-L}
\wedge
(\Box_L \neg \event{N3})_x
\wedge
(\Box_L \neg(\event{Wake}\vee \event{REM}\vee \varphi_{\event{Arousal}}))_x
\Big)
$ \\

\textbf{(14) Interruption logic:} N2 continuation is interrupted by Wake, N3, REM, arousal, or by a major body movement not followed by a characteristic N2 signal. &
$\event{Wake}$, $\event{N3}$, $\event{REM}$, $\event{MajorBodyMove}$, $\event{Spindle}$, $\event{KC}$ &
$\displaystyle
\varphi_{\text{Interrupt}} := 
\event{Wake}
\vee
\event{N3}
\vee
\event{REM}
\vee
\varphi_{\event{Arousal}}
\vee
\Big(
\event{MajorBodyMove}
\wedge
\neg \Diamond_{L/2}\varphi_{\text{Sig}}
\Big)
$ \\

\textbf{(15) Stage N2 scoring:} N2 holds at epoch onset if it is newly initiated by marker-based onset or by transition from N3, or if a prior N2-initiation remains uninterrupted and occupies the majority of the current epoch. &
$\event{Spindle}$, $\event{KC}$, $\event{N3}$, $\event{Wake}$, $\event{REM}$, $\event{MajorBodyMove}$ &
$\displaystyle
\varphi_{\text{N2}} := \exists x\exists u \exists \ell
(\varphi_{\text{N2InitSig}}
\vee
\varphi_{\text{N2InitFromN3}})_x
\vee \Big(
\big(
\varphi_{\text{N2InitG3}}
\vee
\varphi_{\text{N2InitFromN3}}
\big)_{x-u}
\wedge
(\Box_{u+\ell}\neg \varphi_{\text{Interrupt}})_{x-u}
\Big),
\; x,u \in E,\; x \ge u,\; L/2 < \ell \le L
$ \\
\hline
\multicolumn{3}{@{}l}{\textbf{Limb Movements (AASM VII.B)}} \\

\textbf{(16) Respiratory-event macro for LM exclusion:} generic respiratory-event predicate used to exclude limb movements overlapping with respiratory events. &
--- &
$\displaystyle
\pEvent{resp} := \varphi_{\text{apnea}} \vee \varphi_{\text{hypopnea}} \vee \varphi_{\text{RERA}}
$ \\

\textbf{(17) Leg movement (LM):} anterior tibialis EMG increase above threshold lasting 0.5--10\,s, with return below threshold for at least 0.5\,s, excluding respiratory-event overlap/proximity. &
$\event{EMGAboveRestThr}$, $\event{EMGAbove8}$, $\event{EMGAbove2}$ &
$\displaystyle
\pEvent{LM} :=\exists x \exists u \Big(\big(
(\Box_{0.5}\neg \event{EMGAboveRestThr})_{-0.5}
\wedge
(\Box_u \event{EMGAbove8})
\wedge
(\Box_{0.5}\neg \event{EMGAbove2})_{u}
\wedge
\neg \pEvent{resp}
\wedge
\neg \Diamond_{0.5}(\pEvent{resp})
\wedge
\neg \Diamond_{0.5}\big((\pEvent{resp})_{-0.5}\big)\big)_x
\Big),
\; 0.5 \le u \le 10
$ \\

\textbf{(18) Periodic limb movement series (PLMS):} at least four qualifying LMs with inter-movement onset intervals between 5 and 90\,s. &
$-$ &
$\displaystyle
\pEvent{PLMSInit} := \exists x \exists u \exists v \exists w \Big(
(\pEvent{LM})_x
\wedge
\big((\pEvent{LM})_u \wedge \Box_u \neg \pEvent{LM}\big)_x
\wedge
\big((\pEvent{LM})_v \wedge \Box_v \neg \pEvent{LM}\big)_{x+u}
\wedge
\big((\pEvent{LM})_w \wedge \Box_w \neg \pEvent{LM}\big)_{x+u+v}
\Big),
\; 5 \le u \le v \le w \le 90
$ \\
 \\
\end{longtable}
\vspace{-5mm} 
    \parbox{\textwidth}{
         *\scriptsize Representative rules are shown for arousal, apnea/hypopnea, stage N2, and limb movement/PLMS scoring. Rules 1-9, 15, and 17-18 are final scoring specifications; Rules 10-14 and 16 are supporting intermediate definitions or macros used in downstream formulations. We used derived propositions as macros: $\event{Wake}$, $\event{N2}$, $\event{N3}$, and $\event{REM}$ denote staging propositions; $\varphi_{\event{RERA}}$ denotes that a RERA occurs at the reference time; $\varphi_{\event{hypopnea}}$ denotes that a hypopnea occurs at the reference time; $\event{NREM} := \event{N1} \vee \event{N2} \vee \event{N3}$; and $\event{Sleep} := \event{NREM} \vee \event{REM}$.
        $E$ denotes the set of epoch onset times; L denotes epoch length (30 s); $x$ denotes the time variable; $u$, $v$, $\ell$, and $d$ denote nonnegative time-length variables. Specifically, $\varepsilon$ in rule (6) is a clinically configurable constant that captures physiological delay. Above translations are based on AASM Manual version 2.4. Abbreviations: LM, leg movement; PLMS, periodic limb movement series; KC, K-complex; SpO$_2$, peripheral oxygen saturation; REM, rapid eye movement sleep; NREM, non-rapid eye movement sleep; RERA, respiratory effort-related arousal.
    }
}

For the N2 scoring rule, which represents the most complex case, we manually verified that our formulation is consistent with the canonical AASM staging scenarios shown in Figure 7 of Section IV.G of the AASM Manual (p. 23). These examples indicate that the QEL specification successfully captures both same-epoch N2 initiation and the carryover of previously established valid N2 markers, while excluding cases in which a K-complex is followed by an arousal. The formulation also reproduces the continuation and interruption logic specified in the AASM Manual.
\subsubsection*{4.2 \quad Back-Translation Evaluation} 

Across the 12 evaluated final scoring specifications, QEL back-translation achieved consistently high semantic fidelity to the source narrative despite low surface-level textual overlap (Table~\ref{tab:metrics_summary_ci}). Overall, semantic similarity reached 79.3 (95\% CI: 79.0--79.7), whereas lexical sequence matching and ROUGE-L were much lower, at 14.9 (95\% CI: 13.8--16.1) and 18.3 (95\% CI: 17.6--18.9), respectively. This gap indicates that the reconstructed rules often differed from the original wording while still preserving the underlying meaning. It also supports the interpretation that the back-translation was driven primarily by the logical structure of the formulas, rather than by memorization of existing AASM-related knowledge. Notably, even relatively complex scoring rules, such as N2 scoring, remained semantically recoverable from QEL specifications, suggesting that QEL captures the core clinical content of the original rules while abstracting away from their surface textual form.

\section*{5.\quad Discussion}\label{sec:discussion}

In this work, we proposed Ensemble Logic over the rational numbers (QEL), as a sleep-domain formalism for representing the AASM Scoring Manual. QEL allowed us to encode clinically stated timing constraints and event dependencies as precise logical formulas. With identified atomic propositions and temporal constraints identified in the manual, we successfully translated key AASM phenotypes into reviewable and executable QEL specifications. The back-translation findings further supported that QEL preserves clinical meaning even when it abstracts away from the surface wording of the manual.

Translating the natural-language text of the AASM Scoring Manual into machine-executable logic is challenging. The manual’s logical structure is often hidden by dense prose in which formal rules and explanatory notes are semantically entangled. To address this problem, we developed a principled formalization framework that spans the full computational pipeline, from extracting atomic propositions to constructing complex temporal formulas. As validated in our case studies spanning four distinct physiological domains, this formalization distilled verbose guidelines into compact mathematical specifications while revealing and resolving implicit ambiguities.
\Needspace{0.42\textheight}
\begin{wraptable}[15]{l}{0.6\textwidth} 
\centering
\scriptsize 
\captionsetup{font=small,skip=1pt}
\setlength{\tabcolsep}{3pt} 
\renewcommand{\arraystretch}{0.92}
\caption{Semantic and lexical similarity scores by back-translation evaluation. Metrics are presented as mean (95\% CI) across 40 independent runs.}
\label{tab:metrics_summary_ci}
\begin{tabular}{lccc}
\hline
\textbf{Rules} & \textbf{Semantic} & \textbf{Lexical} & \textbf{ROUGE-L} \\
\hline
Stage N2 scoring & 86.1 (85.6-86.6) & 5.2 (4.8-5.6) & 18.3 (17.9-18.7) \\
Arousal & 82.6 (82.6-82.7) & 19.2 (19.0-19.3) & 31.2 (30.8-31.6) \\
Hypopnea 1A & 81.1 (80.9-81.2) & 11.3 (11.1-11.5) & 13.8 (13.5-14.2) \\
Obstructive hypopnea & 80.7 (80.4-81.1) & 6.9 (6.7-7.1) & 14.8 (14.3-15.3) \\
Leg movement (LM) & 79.9 (79.4-80.3) & 9.6 (8.9-10.4) & 23.3 (22.8-23.8) \\
Mixed apnea & 79.5 (78.8-80.1) & 7.0 (6.8-7.1) & 23.8 (22.7-25.0) \\
Obstructive apnea & 80.0 (79.9-80.1) & 43.8 (42.8-44.8) & 22.2 (21.5-22.9) \\
Hypopnea 1B & 79.1 (78.9-79.2) & 12.3 (12.1-12.4) & 10.7 (10.5-11.0) \\
\begin{tabular}{@{}l@{}}Periodic limb movement \\ (PLMS)\end{tabular} & 80.1 (79.8-80.4) & 9.4 (9.0-9.8) & 13.7 (13.2-14.3) \\
Central hypopnea & 75.1 (74.8-75.5) & 7.8 (7.6-8.1) & 12.6 (12.0-13.3) \\
Central apnea & 76.0 (75.7-76.2) & 39.4 (38.2-40.6) & 26.9 (25.8-27.9) \\
Apnea & 71.8 (71.8-71.9) & 7.4 (7.3-7.4) & 7.8 (7.8-7.9) \\
\hline
\textbf{Overall} & \textbf{79.3 (79.0-79.7)} & \textbf{14.9 (13.8-16.1)} & \textbf{18.3 (17.6-18.9)} \\
\hline
\end{tabular}
\vspace{-2pt} 
    \parbox{0.6\textwidth}{\raggedright
        *\scriptsize Values are reported as mean (95\% CI) across 40 independent runs. All metrics are presented on a 0-100 scale, with higher values indicating greater similarity. Semantic denotes embedding-based semantic similarity; Lexical denotes sequence matching ratio; ROUGE-L denotes longest-common-subsequence overlap. The Overall row summarizes results across the 12 final scoring specifications.
    }
\vspace{-4mm} 
\end{wraptable}
Crucially, our translation process went beyond a syntactic mapping of text to logic. It was informed by sleep physiology. A recurring issue is the manual’s conflation of continuous micro-states (e.g., brief wake intrusions) with discrete, epoch-level staging decisions (e.g., N2 scoring over 30-s epochs). We addressed this by using a continuous, dense time domain together with boundary operators parameterized by the epoch length $L$.  Another example is the use of phrases such as “associated with” or “there is” in Hypopnea scoring. A naive, literal translation into logic would strictly interpret these terms as a requirement for the simultaneous occurrence of airflow reduction and physiological consequences (desaturation). However, this ignores physiological latency (e.g., the circulatory delay required for blood oxygen levels to drop.) \cite{hypoxic_burden,younes2004role}. Such an interpretation would yield massive false negatives computationally. By using QEL's metricized modalities, we resolved this semantic gap by introducing an explicit, parameterized latency variable $\epsilon$. Expanding the evaluation window ($\Diamond_{u+\epsilon}$) successfully validated overlapping apneas and subsequent desaturations as contiguous physiological events, directly mirroring expert clinical consensus without requiring arbitrary heuristic relaxations. In addition,  the "majority" axiom exhibits semantic ambiguity by mapping to multiple distinct temporal patterns. Two primary examples include: (i) cumulative duration: requiring the total accumulated time of a fragmented state to exceed 50\% of an epoch, and (ii) sequence truncation, where an interrupting event in the first half of an epoch breaks the continuous persistence of a prior state. In this work, this ambiguity was resolved at the semantic layer by directly evaluating the overlapping intervals with quantifying time terms. Future extensions of QEL will formalize a native majority operator to unify these distinct temporal patterns directly within the logic's syntax.

At the system level, our framework assumed a strict separation between the signal-processing layer and the logical reasoning engine. Sleep observations exhibit strong dependencies, including mutual exclusivity among sleep stages and hierarchical implications induced by physiological thresholds. As a result,  end-to-end logical consistency requires global constraints over the BEST. We therefore argue for a dedicated \emph{Sleep Ontology}, conceptually similar to standardized biomedical vocabularies such as SNOMED CT \cite{bos2006snomed}. This ontology would provide standardized semantic queries and global constraints to the logic layer, ensuring combinations of atomic propositions remain medically consistent.

As an exploratory step toward bridging clinical guidelines with formal methods, this work presents several limitations. First, temporal logics with unrestricted quantifier alternation are generally undecidable. However, the specific QEL fragments utilized in our modeling intentionally rely on bounded time variables and avoid deep quantifier nesting; thus, the practical logic employed here is anticipated to be largely decidable. Second, fully operationalizing this reasoning engine will require substantial engineering efforts to build the aforementioned standardized Sleep Ontology for atomic propositions. Third, QEL does not eliminate interpreter-dependent variability when narrative AASM rules are first formalized, especially where prose leaves clinical terms, temporal associations, or thresholds underspecified (e.g., associated with, ``majority of the epoch''). This limitation is analogous to earlier guideline-representation efforts, in which professional background could influence encoding choices~\cite{ohno-machado1998glif}. QEL mitigates rather than removes this risk by forcing each choice into explicit temporal predicates, Boolean relations, and parameterized bounds, making disagreement auditable. Thus, reproducibility holds by construction after formalization: the same QEL specification, evaluated on the same BEST, yields the same result, while reproducibility of the initial translation requires consensus review and future multi-interpreter validation.

\section*{6. \quad Conclusion}
In this study, we presented a framework that bridges the gap between the natural-language guidelines in the AASM Scoring Manual and machine-executable specifications using Rational Ensemble Logic (QEL). By mapping PSG annotations to Biomedical Event Structure Temporal Models, we translated complex, time-dependent clinical rules into precise mathematical formulas. Our Back-translation experiments further showed that QEL rules can be rendered into clinically interpretable language without substantial loss of meaning. In future work, we will validate these formulations with clinical experts and develop an open-source event library to support standardized, reproducible, and large-scale sleep research.
\paragraph{Acknowledgment.} This work was supported in part by the National Science Foundation (NSF) award IIS2500624,  the National Institutes of Health (NIH) grants (R01NS126690; U24AG098157). The content is solely the responsibility of the authors and does not necessarily represent the official views of the NSF or NIH.

\makeatletter
\renewcommand{\@biblabel}[1]{\hfill #1.}
\makeatother

\bibliographystyle{vancouver}
\small
\bibliography{amia}  

\end{document}